\documentstyle[proceedings,psfig]{crckapb}

\begin{opening}
\title{OLD STELLAR POPULATIONS IN NEARBY DWARF 
GALAXIES$\thefootnote{^1}$}

\author{Eline Tolstoy}
\institute{European Southern Observatory\\
           Garching bei M\"{u}nchen, Germany}
\end{opening}

\runningtitle{Old Stellar Populations...}

\begin{document}
\def\lsim{~\rlap{$<$}{\lower 1.0ex\hbox{$\sim$}}}
\def\gsim{~\rlap{$>$}{\lower 1.0ex\hbox{$\sim$}}}

\begin{abstract}
What can we learn from the somewhat arduous study of old stellar
populations in nearby galaxies? 
Unless the nearby universe is subtly anomalous, it should contain a 
relatively normal selection of galaxies whose histories are representative 
of field galaxies in general throughout the Universe.  We can therefore 
take advantage of our ability to resolve local galaxies into 
individual stars to directly, and accurately, measure star formation 
histories.  The star formation histories are determined from 
numerical models, 
based on stellar evolution tracks, of colour-magnitude diagrams.  
The most accurate information on star formation rates extending back to the 
earliest epoches can be obtained from the structure of the main sequence.  
However, the oldest main sequence turnoffs are very faint, and it is often 
necessary to use the brighter, more evolved, 
populations to infer the star formation 
history at older times. 
A complete star formation history can 
be compared with the spectroscopic properties of galaxies seen 
over a large range of lookback times in redshift surveys.  There 
is considerable evidence that the faint blue galaxies seen in large 
numbers in cosmological surveys are the progenitors of the 
late-type irregular galaxies seen in copious numbers in the Local Group,
and beyond.
We consider how the ``Madau-diagram'', the star formation history
of the Universe, would look if the Local Group were to be
considered representative of the Universe as a whole.

\end{abstract}

\footnotetext[1]{\tt Invited Review, 
to appear in ``Galaxy Evolution: Connecting the Distant
Universe with the Local Fossil Record'', eds. M. Spite, F. Crifo}

\section{Introduction}

The study of resolved stellar populations provides a powerful tool to follow 
galaxy evolution directly in terms of physical parameters such as age (star 
formation history, SFH), chemical composition and enrichment history, initial 
mass function, environment, and dynamical history of the system.  
Photometry of individual stars in at least two filters and the
interpretation of Colour-Magnitude Diagram (CMD) morphology gives the
least ambiguous and most accurate information about variations in
star formation within a galaxy back to the oldest stars.
Some of the physical parameters that affect a CMD are strongly correlated, 
such as metallicity and age, since successive generations of star formation 
may be progressively enriched in the heavier elements. Careful, detailed CMD 
analysis is a proven, uniquely powerful approach 
(e.g.,  Tosi {\it et al.} 1991; Tolstoy \& Saha 1996; 
Aparicio {\it et al.} 1996; Mighell 1997; 
Dohm-Palmer {\it et al.} 1997, 1998;
Hurley-Keller {\it et al.} 1998; Gallagher {\it et al.} 
1998; Tolstoy {\it et al.} 1998) that benefits enormously 
from the high spatial resolution of $HST$ to the point that ground
based CMD analysis is only worthwhile in ideal conditions beyond
about the distance of the Magellanic Clouds.

Because of the tremendous gains in data quality and thus understanding which 
have come from recent high quality CMDs of nearby galaxies it is now 
clearly worthwhile and fundamentally important to complete a survey 
of the resolved stellar populations of all the
galaxies in our Local Group (LG).
This will provide a uniform picture of the global star formation
properties of galaxies with a wide variety of mass, metallicity, gas
content etc. (e.g. Mateo 1998), and
will make a sample that ought to reflect the SFH of the Universe 
and give results which can be compared
to high redshift survey results 
(e.g., Madau {\it et al.} 1998). Initial comparisons suggest these
different approaches do not yield the same results 
(Fukugita {\it et al.} 1998),
but the errors are large.

\begin{figure}
\psfig{file=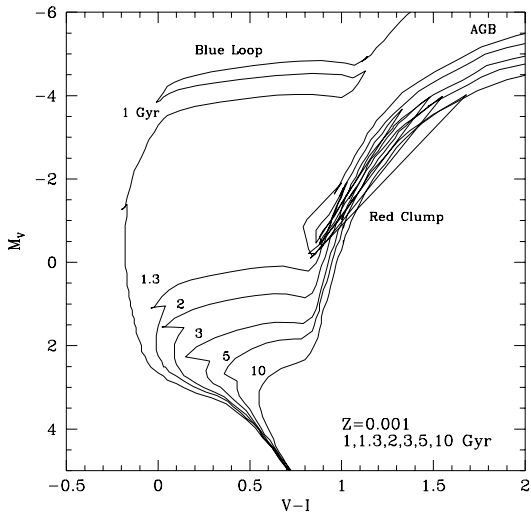,height=10cm,width=10cm}
\vskip-2.5cm
\caption{
Isochrones (Bertelli {\it et al.} 1994) for a single
metallicity (Z=0.001) and a range of ages, as marked in Gyr
at the MSTOs.  Isochrones were designed for single age
globular cluster populations and are best avoided in the
interpretation of composite populations, which can best be modeled
using Monte-Carlo techniques ({\it e.g.} Tolstoy 1996). They are
used here for the purpose of illustration.
}
\vskip-0.5cm
\end{figure}

\section{Colour-Magnitude Diagram Analysis}

Much of our detailed knowledge
of the SFHs of galaxies beyond 1~Gyr ago
comes from the Milky Way and its nearby dSph satellites or from $HST$
CMDs. To date, the limiting factors have been crowding
and resolution limits for accurate stellar photometry from the ground.
$HST$ provides a unique opportunity to extend 
beyond our immediate vicinity and encompass the whole LG.
To date $HST$ has observed the
resolved stellar populations in variety of nearby galaxies
(e.g., dE, NGC~147, Han {\it et al} 1997; Irr, LMC, Geha {\it et al.} 1998;
Spiral, M~31, Holland {\it et al.} 1996; 
BCD, VII~Zw~403, Lynds {\it et al.} 1998; 
dI, Leo~A, Tolstoy {\it et al.} 1998; 
dSph, Leo~I, Gallart {\it et al.} 1998).
For every LG galaxy at
which $HST$ has pointed at we have learnt something new
and fundamentally 
important that was not discernable from ground based images, 
especially in the case of small dIs. The small dIs, like the dSph
appear to exhibit a wide variety of SFHs. These 
results have affected
our understanding of galaxy formation and
evolution by demonstrating the importance of episodic star
formation in nearby low mass galaxies.
The larger galaxies in the LG 
have evidence of sizeable old halos, which appear 
to represent the majority of star formation in the
LG by mass, although the problems distinguishing between effects of age and 
metallicity in a CMD result in a degree of uncertainty in the exact age
distribution in these halos. It is important that
detailed comparative
studies of all galaxies in the LG are made in the future, including the
M~31 and M~33 halo populations, to obtain a picture
of the fossil record of star formation in
galaxies of various types and sizes, and to identify both
commonalities and differences in their SFH across the LG.
In addition to a better understanding of galaxy evolution this
will enable the comparison with cosmological surveys to be made
more accurately.

Stellar evolution theory provides a number of clear predictions, based on 
relatively well understood physics, of features expected in 
CMDs for different age and metallicity stellar 
populations (see Figure~1).  There are a number of clear indicators of 
varying star formation rates ({\it sfr}) at different times which can be 
combined to obtain a very accurate picture of the entire 
SFH of a galaxy.

\noindent{{\it Main Sequence Turnoffs (MSTOs)}}:
If we can obtain deep enough exposures of the resolved stellar populations 
in nearby galaxies we can obtain the {\it unambiguous age information that 
comes from the luminosity of MSTOs}.  Along the Main Sequence itself 
different age populations overlie each other completely making the 
interpretation of the Main Sequence 
luminosity function complex, especially for older 
populations.  However the MSTOs do not overlap each other like this and 
hence provide the most direct, accurate information about the SFH of a 
galaxy.  MSTOs can clearly distinguish between bursting star formation and 
quiescent star formation, ({\it e.g.}  Hurley-Keller {\it et al.} 1998).  

\noindent{{\it The Red Giant Branch (RGB)}}:
The RGB is a very bright evolved phase of stellar evolution, where the star 
is burning H in a shell around its He core.  For a given metallicity the 
RGB red and blue limits are given by the young and old limits 
(respectively) of the stars populating it (for ages $\gsim$1~Gyr).  As a 
stellar population ages the RGB moves to the red, for constant metallicity, 
the blue edge is determined by the age of the oldest stars.  However 
increasing the metallicity of a stellar population will also produce 
exactly the same effect as aging, and also makes the RGB redder.  This is 
the (in)famous age-metallicity degeneracy problem.  The result is that if 
there is metallicity evolution within a galaxy, it impossible to uniquely 
disentangle effects due to age and metallicity on the basis of the optical 
colours of the RGB alone.

\begin{figure}
\vskip0.5cm
\psfig{file=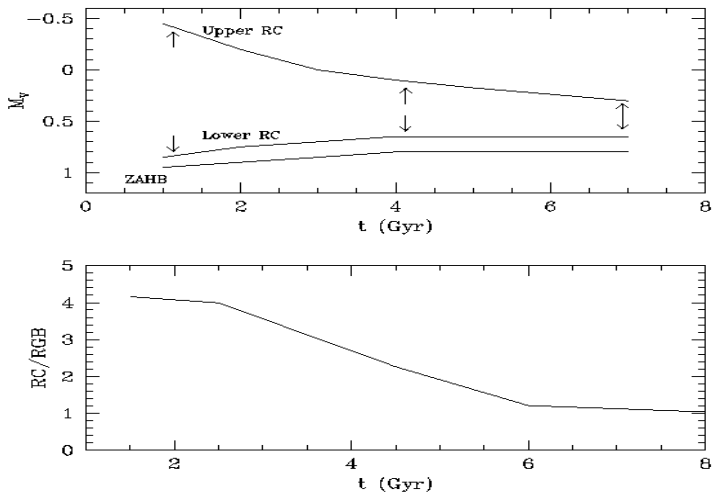,height=7cm,width=10cm}
\caption{
In the top panel are plotted
the results of Caputo, Castellani \& Degl'Innocenti (1995) for the 
variation in the {\it extent} in M$_V$ magnitude of a RC with age, for a 
metallicity of Z=0.0004.  We plot the magnitude of the upper and lower edge 
of the RC versus age, in Gyr.  We can thus clearly see that this extent is 
strong function of the age of the stellar population.  Also plotted is 
M$_V$ of the zero age HB against age.
In the lower panel are plotted the results of running a series of 
Monte-Carlo simulations (Tolstoy 1996) using stellar evolution models at 
Z=0.0004 (Fagotto {\it et al.} 1994) 
and counting the number of RC and RGB stars in the same 
part of the diagram, and thus we determine the expected ratio of RC/RGB 
stars versus age.
}
\vskip-0.5cm
\end{figure}

\noindent{{\it The Red Clump/Horizontal Branch (RC/HB)}}:
Red Clump (RC) stars and their lower mass cousins, Horizontal Branch (HB) 
stars are core helium-burning stars, and their luminosity varies depending 
upon age, metallicity and mass loss (Caputo {\it et al.} 1995).  
The extent in luminosity 
of the RC can be used to estimate the age of the population that produced 
it (Caputo {\it et al.} 1995),  as shown in the upper panel of Figure~2.  
The ratio, t$_{RC}$ / t$_{RGB}$, 
is a decreasing function of the age of the dominant stellar population in a 
galaxy, and the ratio of the numbers of stars in the RC, and the HB to the 
number of RGB is sensitive to the SFH of the 
galaxy (Tolstoy {\it et al.} 1998; Han {\it et al.} 1997).
Thus, the higher the ratio, N(RC)/N(RGB), the younger the dominant stellar 
population in a galaxy, as shown in the lower panel of Figure~2.
This age 
measure is {\it independent of absolute magnitude and hence distance}, and 
indeed these properties can be used to determine an accurate distance 
measure on the basis of the RC (e.g. Cole 1998).
The presence of a large HB population on the other hand (high N(HB)/N(RGB) 
or even N(HB)/N(MS), is caused by a predominantly much older ($>$10~Gyr) 
stellar population in a galaxy.  The HB is the brightest indicator of very 
lowest mass (hence oldest) stellar populations in a galaxy.

\noindent{{\it The Extended Asymptotic Giant Branch (EAGB)}}:
The temperature and colour of the EAGB stars in a galaxy are determined by 
the age and metallicity of the population they represent (see Figure~3).  
However there remain a number of uncertainties in the comparison between 
the models and the data (Gallart {\it et al.} 1994; 
Lynds {\it et al.} 1998).  It is very important that more 
work is done to enable a better calibration of these very bright indicators 
of past star formation events.  In Figure~3 theoretical EAGB isochrones 
(Bertelli {\it et al.} 1994) are overlaid on the HST CMD of a 
post-starburst BCD galaxy VII~Zw403, and we 
can see that a large population of EAGB stars is a bright indicator of a 
past high {\it sfr}, and the luminosity spread depends upon metallicity and 
the age of the {\it sfr}. That the RGB+AGB population of VII~Zw403 
looks so similar to NGC~6822 (Gallart {\it et al.} 1994) is
suggestive that dI and BCD galaxies can easily transform into each
other on very short time scales.

\begin{figure}
\vskip-7cm
\psfig{file=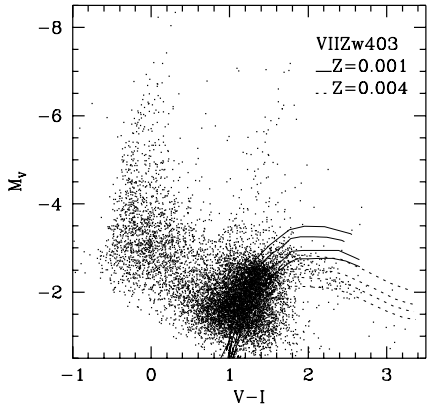,height=14cm,width=14cm}
\vskip-1cm
\caption{
EAGB isochrones (Bertelli {\it et al.} 1994)
for metallicities, Z=0.001 and Z=0.004, are shown superposed on the 
observed CMD of VII~Zw403 (Lynds {\it et al.} 1998).  
For each metallicity the isochrones 
are for populations of ages 1.3, 2, 3, and 5 Gyrs, with the youngest 
isochrone being the brightest.  This shows the potential discriminant 
between the age and metallicity of older populations, if the models could 
be better calibrated to a known SFH, {\it e.g.}  for a nearby EAGB rich system 
like NGC~6822 where old MSTOs are observable.
}
\vskip-0.5cm
\end{figure}

\section{The Connection to High Redshift}

Star-forming, dI galaxies represent the largest fraction by number of 
galaxies in the LG, and it is clear from deep imaging surveys that this 
number count dominance appears to {\it increase} throughout the Universe 
with lookback time (Ellis 1997).  The large numbers of ``Faint Blue 
Galaxies'' (FBG) found in deep imaging-redshift surveys appear to be 
predominantly intermediate redshift ($z<1$, or a look-back time out to 
roughly half a Hubble time), intrinsically {\it small} late type galaxies, 
undergoing strong bursts of star 
formation (Babul \& Ferguson 1996).  Thus we can assume 
that the dIs we see in the LG are a 
cosmologically important population of galaxies which can be used to trace 
the evolutionary changes in the {\it sfr} of the Universe with redshift.  
The ``Madau-diagram'' (Madau {\it et al.} 1998) 
uses the results of redshift surveys to 
plot the SFH of the Universe against redshift.  It predicts that most of 
the stars that have formed in the Universe have done so at redshifts, {\it 
z} $\sim 1 - 2$.  If it is correct, then the MSTOs from the most active 
period of star formation in the Universe will be easily visible as 
7$-$9~Gyr old MSTOs in the galaxies of the LG (e.g. Rich 1998).
Determining accurate SFHs 
for all the galaxies in the local Universe using CMD analysis provides an 
alternate route to and thus check upon the Madau-diagram.

Recent detailed CMDs of several nearby galaxies and self-consistent grids 
of theoretical stellar evolution models have transformed our understanding 
of galactic SFHs.  Most of the dI CMDs to date suggest
that the {\it sfr} was higher in 
the past, although the peak in the {\it sfr} has 
occured at relatively recent 
times as defined by Madau-diagram (the peaks occur at z=0.1$-$0.2, within the 
first bin).  The Mateo review of {\it all} LG dwarf galaxies (Mateo 1998)
and studies of M31 and our Galaxy (Renzini 1998), 
on the other hand, suggest that the LG 
had its most significant peak in star formation $>$10~Gyr ago (i.e at 
z~$>$~3), the epoch of halo formation.  Many galaxies contain large
numbers of RR~Lyr variables (or HB) and/or globular clusters which can only 
come from a
significant older population.  It is possible that dI galaxies have quite 
different SFHs to the more massive galaxies.  Thus although the small dI 
galaxies in the LG have been having short, often intense, bursts of star 
formation in comparatively recent times this is not representative of the 
majority of the star formation in
the LG. However direct observations of 
the details of the oldest star forming episodes in any galaxy
are limited at best.
This is an area where advanced CMD analysis 
techniques have been developed (e.g. Tolstoy \& Saha 1996) 
and telescopes with sufficient 
image quality exist and the required deep, high quality imaging are 
observations are waiting to be made.

\begin{figure}
\psfig{file=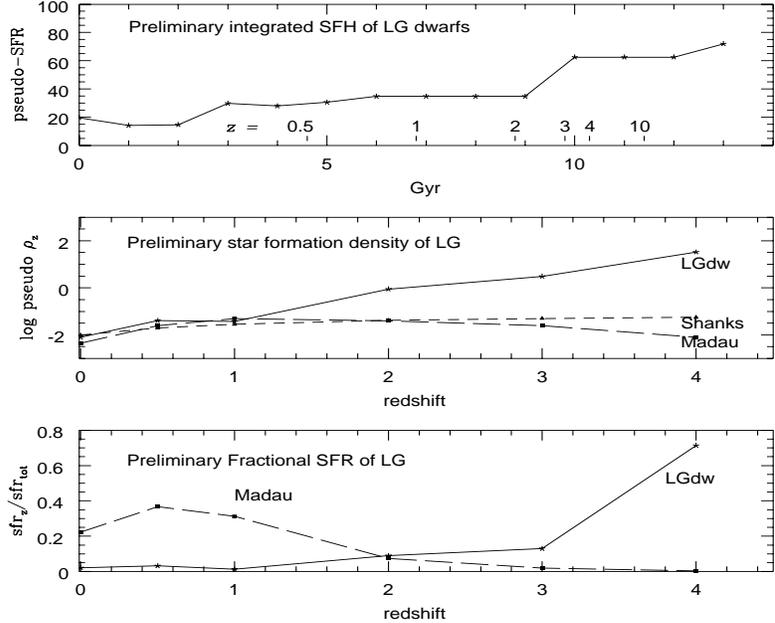,height=9cm,width=11cm}
\caption{
In the upper panel is
a {\it rough} summation of the {\it sfr}s of the LG dwarf galaxies with 
time (data taken from Mateo 1998) to obtain the integrated SFH of 
all the LG dwarfs.  The redshifts corresponding to lookback times (for H$_0 
= 50$, q$_0 = 0.5$).  In the middle panel, a wild extrapolation is made; 
the assumption that the integrated SFH of the LG {\it dwarfs} in the upper 
panel is representative of the Universe as a whole.  The resulting star 
formation density of the LG versus redshift is plotted using the same 
scheme as Madau {\it et al.} (1998) and Shanks {\it et al.} (1998), and these 
two models are also plotted and the LG curve is {\bf arbitrarily}, and with 
a very high degree of uncertainty, normalised to the other two models.  In 
the lowest panel the The LG dwarf {\it sfr} as a fraction of the total star 
formation integrated over all time is plotted versus redshift, and the 
Madau curve is also replotted in this form, for the volume of the LG.  This 
highlights the totally different distribution of star formation with 
redshift found from galaxy redshift surveys and what we appear to 
observe in the stellar population of the LG.
}
\vskip-0.5cm
\end{figure}

Figure~4 summarizes what can currently be said
about the SFH of the LG and how 
this compares with the Madau {\it et al.} (1998) and Shanks {\it et 
al.} (1998) redshift survey predictions.  We have not included the 
dominant large galaxies in the LG, the Galaxy and M~31, 
but the SFH of the combined dwarfs is broadly consistent with what is known 
about the SFH of these large systems.  They have, as far as we can tell, had 
a global {\it sfr} that has been gradually but steadily declining since 
their (presumed) formation epoch $>$10~Gyr ago.  There is currently no 
evidence for a particular peak in {\it sfr} around 7$-$9~Gyr ago or any 
other time, as predicted by the Madau-diagram for either large galaxies
or dwarfs. The dominant population by mass in the LG dwarfs are dE,
if dIs are singled out a population with a star formation
peak in the Madau-diagram range can be found. But at present the statistics
are too limited to determine the typical fraction of old population
in LG dIs. There is 
clearly a total mismatch between the SFH of the LG and the results 
from the redshifts surveys.  This might hint at serious incompleteness 
problems in high redshift galaxy surveys, which appear to miss passively 
evolving systems in favour of small bursting systems.

The recent HST CMD results 
give much cause for optimism that we can hope to sort 
out in detail the SFH of all the different types of galaxies within in the 
LG if only HST would point at them occasionally.  There is also great 
potential for ground based imaging using high quality imaging telescopes 
with large collecting areas, such as VLT is clearly going to be.

\def\aj{{\it Astron.~J.}}			
\def\araa{{\it Ann.~Rev.~Astron.\&~Astrophys.}}		
\def\apj{{\it Astroph.~J.}}			
\def\apjl{{\it Astroph.~J.~Lett.}}		
\def\apjs{{\it Astroph.~J.~Supp.}}
\def\astap{{\it Astron. \& Astrophys.}}
\def\astaps{{\it Astron. \& Astrophys. Supp.}}

\end{document}